\documentclass{article}
\usepackage{latexsym,amsfonts,amssymb,amsmath,amsthm}
\usepackage{graphicx}

\begin{document}

\title{Coordinate time and proper time in the GPS}
\author{T.~Matolcsi\thanks
{Department of Applied Analysis, E\"otv\"os University,
P\'azm\'any P. s\'et\'any 1C., H--1117 Budapest, Hungary,
matolcsi@cs.elte.hu, supported by OTKA-T 048489.},~
M.~Matolcsi\thanks {Alfr\'ed R\'enyi Institute of Mathematics,
Hungarian Academy of Sciences, Re\'altanoda utca 13--15., H--1053,
Budapest, Hungary, matomate@renyi.hu, (also at BME, Dept. of
Analysis, Egry J. u. 1, Budapest, Hungary) supported by OTKA-PF
64061, T 049301, T 047276.}} \maketitle

\newcommand\p{\mathbf p}
\newcommand\piu{\pi_\uu}
\newcommand\e{\mathbf e} \newcommand\n{\mathbf n}
\newcommand\w{\mathbf w}
\newcommand\om{\omega}
\newcommand\rr{\mathbb R}
\newcommand\bs{\mathbf b}
\newcommand\veg{\end{equation}}

\newcommand\uu{\mathbf u} \newcommand\g{\mathbf g}
\newcommand\Hh{\mathbf H} \newcommand\D{\mathrm D}
\newcommand\Eu{\mathbf E_\uu} \newcommand\N{\mathbf N}
\newcommand\h{\mathbf h} \newcommand\1{\boldsymbol 1}
\newcommand\x{\mathbf x}
\newcommand\y{\mathbf y}
\newcommand\I{\mathbf I} \newcommand\R{\mathbf R}
\newcommand\vv{\mathbf v}
\newcommand\M{\mathbf M}
\newcommand\V{\mathbf V}
\renewcommand\t{\mathbf t}
\newcommand\Pp{\mathbf P}
\newcommand\s{\mathbf s}
\newcommand\si{\sigma}
\newcommand\be{\begin{equation}}
\newcommand\U{\mathbf U}
\newcommand\q{\mathbf q}
\newcommand\Om{\boldsymbol\Omega}
\newcommand\E{\mathbf E}
\newcommand\z{\mathbf z} \renewcommand\L{\mathbf L}
\newcommand\zz{\hat{\mathbf z}}
\newcommand\A{\mathbf A}
\newcommand\uc{\uu_c}
\newcommand\as{\mathbf a}
\newcommand\al{\alpha}
\newcommand\hal{\hat\alpha}
\newcommand\bb{\beta}
\newcommand\hbb{\hat\beta}
\newcommand\dd{\mathbf d}
\newcommand\kk{\mathbf k} \newcommand\hh{\mathbf h}
\newcommand\RR{\mathbb R}
\newcommand\F{\mathbf F}
\newcommand\De{\Delta}
\newcommand\se{s_E}

\begin{abstract}

The Global Positioning System (GPS) provides an excellent
educational example as to how the theory of general relativity is
put into practice and becomes part of our everyday life. This
paper gives a short and instructive derivation of an important
formula used in the GPS, and is aimed at graduate students and
general physicists.

The theoretical background of the GPS (see \cite{ashby}) uses the
Schwarzschild spacetime to deduce
 the {\it approximate} formula ,  $ds/dt\approx
1+V-\frac{|\vv|^2}{2}$, for the relation between the proper time
rate $s$ of a satellite clock and the coordinate time rate $t$.
Here $V$ is the gravitational potential at the position of the
satellite and $\vv$ is its velocity (with light-speed being
normalized as $c=1$). In this note we give a different derivation
of this formula, {\it without using approximations}, to arrive at
$ds/dt=\sqrt{1+2V-|\vv|^2 -\frac{2V}{1+2V}(\n\cdot\vv)^2}$, where
$\n$ is the normal vector pointing outward from the center of
Earth to the satellite. In particular, if the satellite moves
along a circular orbit then the formula simplifies to
$ds/dt=\sqrt{1+2V-|\vv|^2}$.

We emphasize that this derivation is useful mainly for educational
purposes, as the approximation above is already satisfactory in
practice.
\end{abstract}

\section{Introduction}

The most significant application  of the theory of General
Relativity in everyday life, arguably, is the Global Positioning
System. The GPS uses accurate, stable atomic clocks in satellites
and on the ground to provide world-wide position and time
determination. These clocks have relativistic frequency shifts
which need to be carefully accounted for, in order to achieve
synchronization in an underlying Earth-centered inertial frame,
upon which the whole system is based. For educational purposes it
is very instructive to see how a highly abstract theory such as
the general theory of relativity becomes a part of our everyday
life.

Throughout this note we will normalize universal constants for
simplicity, so that light speed and the gravitational constant are
the unity, $c=1$, $\gamma=1$.

Let us briefly sketch here how the GPS works, including the major
ingredients for the reader's convenience (a full and detailed
description is available in \cite{ashby}). In order to determine
your position at the surface of Earth you must be able to see some
(at least four) GPS-satellites simultaneously, and know their
position and your distance to them (with this much information at
hand, it is then elementary geometry to determine your position).
To make this possible, every GPS-satellite emits signals
continuously, describing its position and its local time
$t_{sat}$. If your GPS-device measures its own local time,
$t_{dev}$, then you can use the constancy of the speed of light
$c$ to calculate your distance $d$ to the satellite as
\begin{equation}\label{basic}d=(t_{dev}-t_{sat})c=t_{dev}-t_{sat}
\end{equation}
 keeping in mind the
normalization $c=1$. (We remark here that in practice the
GPS-device is unable to measure time with sufficient accuracy,
therefore $t_{dev}$ also has to be deduced from the data sent by
the satellites. However, this detail is not important for the
purposes of this paper.)

This looks simple enough, but the problem is that formula
\eqref{basic} is only valid if all clocks are {\it synchronized}
in some special relativistic {\it inertial frame}. Therefore, for
the purposes of the GPS one imagines that an inertial frame is
attached to the center of Earth, and we try to synchronize all
clocks such that they measure the time $t$ of this ideal inertial
frame. This means that one should imagine 'ideal' clocks placed
everywhere in the vicinity of Earth, measuring the time $t$, and
thus $t_{sat}$ and $t_{dev}$ should be the read-outs of these
ideal clocks at the place of the satellite when the signal
originates and at the place of your device when the signal is
received. However, what we {\it actually can} measure is the
proper time $t_E$ of stationary clocks on the surface of Earth
(which is measured in time-keeping centers throughout the world),
and the proper time $s$ of satellite clocks. Therefore, we need to
establish a relation between the time rates $t_E$, $s$ and $t$.

Fortunately, the time rate $t_E$ measured on Earth is independent
of where you are (i.e. the same manufactured clocks beat the same
rate in London or Tokyo or New York) and differs from the 'ideal'
time rate $t$ only by a multiplicative constant,
\begin{equation}\label{szokas2}
\frac{dt}{dt_E}=1-\Phi_0,
\end{equation}
where $\Phi_0$ is a constant corresponding to the Earth's geoid
(see \cite{ashby}). This relation is very convenient because the
ideal time rate $t$ can be replaced by $t_E$, something we can
actually measure, and an equation of the form \eqref{basic} still
remains valid after rescaling by the factor $1-\Phi_0$. This
leaves us with the task of establishing a relation between $s$ and
$t$ (or $s$ and $t_E$, whichever turns out to be more convenient).

To determine the relation of the proper time rate $s$ of the
satellite clocks and the time rate $t$ of ideal clocks measuring
the coordinate-time of the underlying Earth-centered inertial
frame, the customary theoretical framework \cite{ashby} is to use
Schwarzschild spacetime, and arrive at the formula
\be\label{szokas} ds/dt\approx 1+V-\frac{|\vv|^2}{2}, \veg after
several first-order approximations in the calculations. Here $V$
denotes the gravitational potential at the position of the
satellite, and $\vv$ is the velocity of the satellite measured in
the underlying non-rotating Earth-centered inertial frame. Formula
\eqref{szokas} is the internationally accepted standard relating
the clock frequencies, as described in \cite{ashby, kouba} and
references therein.

We remark here that the derivation of formula \eqref{szokas2} is
somewhat more involved than that of formula \eqref{szokas}. In
fact, for the purposes of deriving \eqref{szokas2} one needs to
{\it modify} the Schwarzschild metric by a small term, taking into
account multipole contributions corresponding to the Earth's
geoid. However, for formula \eqref{szokas} the modifying term is
{\it disregarded} and the standard Schwarzschild metric is used,
the argument being that the modifying term becomes negligible far
enough from the Earth surface, where the satellites orbit (cf.
\cite{ashby}). In this paper we do not include the derivation of
formula \eqref{szokas2} and the modifying term which is used
there.

In general, in physics it is justified to use approximate formulae
for two different reasons. One is that in some cases the
derivation of an exact result is analytically not possible. The
other is that in certain cases an exact analytic derivation, even
if it exists, would lead to involved and lengthy calculations thus
concealing the important and possibly very simple aspects at the
heart of the issue. In calculations involving the theory of
relativity (in particular, general relativity) there is a tendency
to turn to approximations {\it automatically}, due to the involved
nature of the theory. However, in some rare cases a modified point
of view and an adequate choice of coordinate system can lead to
exact results.

In this note, {\it we adhere to the standard theoretical
framework} used in \cite{ashby, kouba}, but we point out that an
{\it exact formula can be derived} in a very simple manner for the
clock frequency rate \eqref{szokas} in question. Instead of using
the customary isotropic coordinates we will treat Schwarzschild
spacetime with Schwarzschild coordinates (an entirely
coordinate-free derivation of the same formulae can be found in
\cite{eredeti} but it is rather cumbersome). This treatment of
Schwarzschild spacetime is motivated by a similar account of
special relativity in \cite{matolcsi} and that of general
relativity in \cite{sachs}. The coordinate-free point of view
often has the advantage of conceptual clarity and, in this
particular case, brevity of calculations.

Our results here are mostly of educational and theoretical
interest as the existing formula \eqref{szokas} provides good
approximation to the desired precision in the GPS (see
\cite{kouba}).

\section{Schwarzschild's spacetime}

Schwarzschild's spacetime describes the gravitational field of a
pointlike inertial mass $m$. It is a well-known model of general
relativity, but we include its short description here for
convenience.

Let us introduce some notation. Let $E$ be a three-dimensional
Euclidean vector space, the inner product of $\x , \y \in E$ being
denoted by $\x \cdot \y$. For $0\neq \x\in E$ we put
\be\n(\x):=\frac{\x}{|\x|}
\end{equation} for the outward normal vector at $\x$.

Consider $\rr\times E$  as a spacetime manifold with its usual
special relativistic metric, i.e. the Lorentz form given by
\be\label{lorentz}
\begin{pmatrix} -1 & 0 \\
0 & I\end{pmatrix},
\end{equation} where $I$ is the identity matrix. For a world point $(t,\x)\in\rr\times E$ one should think of $t$ as the
synchronization time corresponding to the center of Earth, and
$\x$ as the space vector pointing to the world point from the
center of Earth. The Lorentz form \eqref{lorentz} means that the
Lorentz length-square of a four-vector $(s, \q)$ is given by the
usual formula $|(s, \q)|^2_{Lor}=-s^2+|\q |^2$. This special
relativistic spacetime will be called the Earth Centered Reference
Frame (ECRF). This is an 'ideal' special relativistic frame, and
the task in the GPS is to achieve synchronization of clocks in
this underlying inertial frame.

Introduce the potential $V(\x):=-\frac{m}{|\x|}$ on $E$, and
restrict your consideration to world points $(t,\x)\in \rr\times
E$ for which $1+2V(\x)>0$ (i.e. for world points outside the
Schwarzschild radius). Then it is easy to see (cf. \cite{bergmann,
pauli, weinberg}) that for such world points, the standard form of
the Schwarzschild metric in $\rr\times E$ (i.e. a smooth
collection of Lorentz forms $g(t,\x)$ depending on the world
points) takes the form: \be\label{schw}
g(t,\x)=\begin{pmatrix} -(1+  2V(\x)) & 0 \\
0 & I -  \frac{2V(\x)}{1+2V(\x)}\n(\x)\otimes\n(\x).\end{pmatrix}
\end{equation}

In particular, when we measure the length-square of any {\it space
vector} $\q$ at the space point $\x$ in Schwarzschild's metric we
obtain \be\label{slength}|\q|^2_{Sch}:= |\q|^2 -
\frac{2V(\x)}{1+2V(\x)}\bigl(\n(\x)\cdot\q)\bigr)^2.\end{equation}
Similarly, when we measure the Schwarzschild length-square of a
{\it four-vector} (e.g. a four-velocity) $(s,\q)$ at the point
$\x$ we get \be\label{slength2}|(s, \q)|^2_{Sch}=
-\bigl(1+2V(\x))\bigr) s^2 +|\q|^2_{Sch}.\end{equation}

\section{A satellite in Earth's gravitational field}\label{sat}

In this section we derive the formula relating the proper time $s$
of satellite vehicle-clocks to the ideal time $t$ of the
underlying inertial frame.

A material point in spacetime is described by a world line
function
\be \rr\to\rr\times E, \qquad s\mapsto
\bigl(t(s),x(s)\bigr)
\end{equation}
where $s$ is the proper time of the material point. The
four-velocity, $\bigl(\dot t(s),\dot x(s)\bigr),$ of the material
point always satisfies \be |\bigl(\dot t(s),\dot
x(s)\bigr)|^2_{Sch}=-1.
\end{equation}
Therefore, using equation \eqref{slength2} we obtain
\be\label{satel}-\bigl(1+2V(x(s))\bigr)\dot t(s)^2 + |\dot
x(s)|_{Sch}^2 = -1\end{equation}

The time function $t(s)$ can be inverted, with $s(t)$ denoting the
proper time instant $s$ corresponding to the ECRF-time $t$. Let
$\vv(t):=\frac{dx(t)}{dt}$ denote the relative velocity of the
material point with respect to the center of the Earth. Note that
by the chain rule we have $\vv(t)=\dot x(s(t))\frac{ds}{dt}$. Now,
rearranging \eqref{satel} we obtain \be \frac{\left
(\frac{ds}{dt}\right )^2+|\dot x(s(t))|_{Sch}^2\left
(\frac{ds}{dt}\right )^2}{1+2V(x(s(t)))}=1
\end{equation}
which yields \be
\frac{ds}{dt}=\sqrt{1+2V(x(s(t)))-|\vv(t)|_{Sch}^2}
\end{equation}

Finally, applying \eqref{slength} we obtain the desired relation
\be \label{precise}\frac{ds}{dt}=\sqrt{1+2V-|\vv|^2
-\frac{2V}{1+2V}(\n\cdot\vv)^2}
\end{equation}

Using a series expansion, assuming that all the terms on the
right-hand side are much less than 1, we get back the {\it
approximate} formula \eqref{szokas}
\be\label{32}\frac{ds}{dt}\approx 1 + V -\frac{|\vv|^2}{2}.\veg

Of course, there is still some work to be done before one can
apply equation \eqref{basic} in the GPS. Instead of
$\frac{ds}{dt}$ what we really need is the function $t(s)$,
because the signal emitted by the satellite contains the proper
time instant $s$ measured by clocks on board, and we would like to
replace it by the coordinate-time instant $t$ which it corresponds
to. However, having deduced formula \eqref{precise} (or its
approximation \eqref{szokas}) it is possible to obtain the
function $t(s)$, or at least a good enough approximation of it.
This is described very well in full detail in \cite{ashby}, and we
feel it inappropriate to repeat those calculations word-by-word
here. Nevertheless, we warmly recommend that the reader turn to
\cite{ashby} for the interesting details.


\begin{thebibliography}{1}

\bibitem{ashby}
N. Ashby: Relativity in the global positioning system, in   {\it
100 years of relativity},  257--289, World Sci. Publ., Hackensack,
NJ, (2005). (also at Livingreviews in Relativity,
http://relativity.livingreviews.org/Articles/lrr-2003-1/)

\bibitem{bergmann}
P.G. Bergmann, {\it Introduction to the theory of relativity},
Dover, New York (1978), ch.XIII, pg.203, eq(13.25).

\bibitem{kouba} J. Kouba: Improved relativistic
transformations in GPS, {\it GPS Solutions} (2004), {\bf 8},
170-180.

\bibitem{matolcsi} T. Matolcsi:  {\it Spacetime without Reference Frames}, Akad\'emiai
Kiad\'o Budapest, 1993.

\bibitem{eredeti} T. Matolcsi, M. Matolcsi: GPS revisited: the relation of proper time and coordinate time,
{\em arXiv:math-ph/0611086v1}

\bibitem{pauli}
W. Pauli, {\it Theory of relativity}, Dover, New York (1981),
pg.166, eq(421a).

\bibitem{sachs} R. K. Sachs, H. Wu: {\it General Relativity for Mathematicians},
Springer, New York, (1977).

\bibitem{weinberg} S. Weinberg, {\it Gravitation and Cosmology}, J. Wiley \& Sons Inc., New York,
(1972), pg.181.

\end{thebibliography}
\end{document}